\documentclass[conference]{IEEEtran}
\IEEEoverridecommandlockouts
\usepackage{cite}
\usepackage{amsmath,amssymb,amsfonts}
\usepackage{amsthm}
\usepackage{algorithmic}
\usepackage{graphicx}
\usepackage{textcomp}
\usepackage{xcolor}
\usepackage{xurl} % added by me
\usepackage{hyperref} % added by me
\hypersetup{pdfborder = 0 0 0}

\def\BibTeX{{\rm B\kern-.05em{\sc i\kern-.025em b}\kern-.08em
    T\kern-.1667em\lower.7ex\hbox{E}\kern-.125emX}}

\newtheorem{theorem}{Theorem}
\newtheorem{lemma}[theorem]{Lemma}

\begin{document}

\title{IPv6 Bitcoin-Certified Addresses\\
}

\author{\IEEEauthorblockN{Mathieu Ducroux}
\IEEEauthorblockA{\textit{nChain AG} \\
Zug, Switzerland \\
m.ducroux@nchain.com}
}

\maketitle

\begin{abstract}
A pivotal feature of IPv6 is its plug-and-play capability that enables hosts to integrate seamlessly into networks. In the absence of a trusted authority or security infrastructure, the challenge for hosts is generating their own address and verifying ownership of others. Cryptographically Generated Addresses (CGA) solves this problem by binding IPv6 addresses to hosts’ public keys to prove address ownership. CGA generation involves solving a cryptographic puzzle similar to Bitcoin’s Proof-of-Work (PoW) to deter address spoofing. Unfortunately, solving the puzzle often causes undesirable address generation delays, which has hindered the adoption of CGA. In this paper, we present Bitcoin-Certified Addresses (BCA), a new technique to bind IPv6 addresses to hosts’ public keys. BCA reduces the computational cost of generating addresses by using the PoW computed by Bitcoin nodes to secure the binding. Compared to CGA, BCA provides better protection against spoofing attacks and improves the privacy of hosts. Due to the decentralized nature of the Bitcoin network, BCA avoids reliance on a trusted authority, similar to CGA. BCA shows how the PoW computed by Bitcoin nodes can be reused, which saves costs for hosts and makes Bitcoin mining more efficient.
\end{abstract}


\section{Introduction}
As IPv6 adoption is gaining significant momentum, it has become critical to mitigate attacks such as those targeting the Neighbor Discovery Protocol (NDP) which provides link-layer address resolution \cite{SimNDP}. NDP operates under the premise that the network consists of trusted hosts. However, this assumption does not always hold, especially in public wireless networks where minimal authentication mechanism is required to join the link. This gives the opportunity for an attacker to spoof the addresses of legitimate hosts and launch Denial-of-Service (DoS), Man-in-the-middle, and other network-related attacks \cite{NikNDPAttacks}. To overcome these issues, the SEcure Neighbor Discovery (SEND) protocol has been introduced \cite{KemSEND}. SEND employs Cryptographically Generated Addresses (CGA) to provide authentication of IPv6 addresses without relying on a trusted authority or additional security infrastructure \cite{AurCGA}. CGA has proven to be useful in other environments and its usage has been proposed to secure the Shim6 multihoming protocol \cite{NoeShim6} and the Mobile IPv6 protocol \cite{ArkMoIPv6}.

A CGA is an IPv6 address whose interface identifier (its rightmost 64 bits) is generated by hashing a public key and auxiliary parameters. Any host can verify the ownership of an address by recomputing the address from the public key and requesting a signature from the corresponding public key. To reinforce the binding between the public key and the address, CGA introduced the hash extension technique to the address generation process \cite{Aur2CGA}. This technique requires hosts to solve a partial hash inversion puzzle, similar to Proof-of-Work (PoW)-based systems \cite{JakPoW}. The puzzle solution is hashed together with the public key to generate the address. The difficulty of the puzzle is chosen by hosts depending on their computational power. Increasing the difficulty of the puzzle increases the resistance of an address against spoofing attacks. On the other hand, it also increases the address generation time.

The issue with CGA is that it trades security for performance without being able to offer a good balance between the two. As noted in the original CGA RFC, the hash extension technique is effective if the computational power of attackers and hosts grow at the same rate \cite{AurCGA}. In reality, attackers benefit from a linear increase in attack speed by investing in parallel hardware. This leaves hosts with limited parallel hardware highly susceptible to spoofing attacks. For standard devices, it has been shown that the cost of generating an address with high security can be prohibitive \cite{AlSTBCGA, BosCGA++, AlsSEND}. The issue becomes particularly problematic on mobile networks, in which devices have limited computational power and operations such as handovers must be completed within a few milliseconds \cite{QadCGAMobile}. Additionally, the high computational cost of CGA generation disincentivizes hosts from changing their address frequently, which exposes them to privacy-related attacks \cite{AurCGA}. Using the same interface identifier for a long period of time makes it possible for an attacker to monitor and correlate the activity of a host, even as it changes subnet. The correlation can be done on the characteristics of the intercepted packets, such as size or timing \cite{GonSLAAC}.

Several approaches have been proposed to solve the performance issue of CGA. First, a time-based termination condition can be added to the address generation process \cite{AlSTBCGA}. This ensures that the address generation time does not exceed a predefined value. However, the trade-off between security and performance remains the same and hosts with limited computational power remain highly susceptible to spoofing attacks. Another suggestion consists of performing the work in advance or offline, rather than in real time when a new address is needed \cite{AurCGA}. Even though this might reduce the delay needed to obtain a new address, it does not reduce the computational burden of generating addresses. To guarantee a good level of security while maintaining a reasonable computational cost, a local key server can be employed \cite{GuaQuickCGA}. The server is responsible for performing the work in advance and serving keys to hosts that join the network. However, this model introduces a single point of failure and involves setting up additional security infrastructure, which is against the original CGA design \cite{Aur2CGA}.

The work performed by hosts in CGA exhibits similarities to that performed by Bitcoin nodes to secure the Bitcoin blockchain. Launched in 2009, Bitcoin is the first and most successful implementation of a blockchain \cite{NakBit}. Bitcoin nodes consists of a distributed and decentralized network of nodes which agrees on the next block of transactions to be appended to the blockchain through PoW \cite{JakPoW}. In PoW, Bitcoin nodes compete to solve a partial hash inversion puzzle tied to a block. The first node to complete the puzzle obtains the right to add the block to the blockchain and claim the associated reward. The competitive nature of PoW coupled with the development of highly performant hardware and the increasing popularity of Bitcoin has led to a steady increase of the total hash rate of the Bitcoin network. In February 2023, it peaked at about $300 \times 10^{18}$ hashes per second \cite{BitHashrate}, making Bitcoin the most powerful distributed hashing system in the world.

In this paper, we introduce Bitcoin-Certified Addresses (BCA), a new technique to generate IPv6 addresses from a host’s public key registered on the Bitcoin blockchain. The PoW computed by Bitcoin nodes is used to secure the binding between the public key and the address, thereby making the computational cost of address generation minimal for hosts. Compared to CGA, the security of the binding is improved since it is guaranteed by Bitcoin nodes instead of hosts’ devices which typically have limited computational power. The reward mechanism in Bitcoin incentivizes nodes to invest in the latest hardware technology, as evidenced by the network’s increasing hash rate \cite{BitHashrate}. As a result, the growth in computational power of Bitcoin nodes is likely to be aligned with that of a powerful attacker. In CGA, this assumption is not true because standard devices tend to have limited hardware available. This renders them vulnerable to attackers that are able to invest in lots of parallel hardware.

BCA improves on previous proposals which suggested delegating the expensive work performed during address generation to external computers \cite{AurCGA, GuaQuickCGA}. First, BCA is more efficient as it does not require setting up extra infrastructure. Instead, it relies on the existing Bitcoin network and the already expended work of the Bitcoin nodes. Secondly, the highly distributed nature of the Bitcoin network makes it resistant to DoS attacks and manipulation by an attacker. Finally, the financial incentives for performing the work are clearer. BCA provides atomic payments in the form of transaction fees to remunerate Bitcoin nodes for their work.

BCA demonstrates how the PoW computed by Bitcoin nodes can be reused, which makes Bitcoin mining more efficient. A common criticism of Bitcoin mining is that it wastes computational power and energy without having any intrinsic value \cite{SalPoS, BecPoW}. Several projects proposed to replace PoW with computational problems that have direct real-world applications \cite{MilPermacoin, ChaHybridMining, KinPrimecoin}. Some blockchains make additional value of the PoW computed in Bitcoin by reusing it in their consensus protocol, a technique referred to as merge-mining \cite{JudMergedMining, TasBabylon, Namecoin}. BCA shows that besides securing blockchains, the PoW computed by Bitcoin nodes can be used to secure other protocols that would otherwise require users to expend a lot of computational resources themselves.

Our contributions in this paper are the following:

\begin{itemize}
\item We introduce BCA, a new technique to generate IPv6 addresses from public keys. The technique is efficient and provides a good level of security for all hosts, regardless of their computational power.
\item We present a detailed analysis of BCA and CGA. In particular, we analyse the cost of spoofing attacks in both techniques. We show that in practice, BCA provides better security against spoofing attacks than CGA. 
\item We provide an implementation of BCA to demonstrate its usability. We also provide an implementation of CGA and evaluate the two techniques. Our evaluation shows that to obtain the same level of security as BCA, generating an address in CGA takes an average of 696.15 s, while it takes only 0.00096 ms in BCA on a standard laptop.
\end{itemize}

The rest of the paper is organized as follows. Section \ref{section:cga} presents CGA and analyzes its resistance against spoofing attacks and costs. Section \ref{section:bitcoin} describes the Bitcoin protocol. Section \ref{section:bca} introduces the proposed BCA technique and gives a detailed analysis of it. Section \ref{section:implementation} describes our BCA implementation and compares its performances to that of CGA. Finally, the conclusions are presented in Section \ref{section:conclusion}.

\section{Cryptographically Generated Addresses (CGA)} \label{section:cga}

In this section, we describe the CGA technique and analyze its resistance against spoofing attacks and its costs.

\subsection{CGA Specification}

The objective of CGA is to prevent spoofing of IPv6 addresses by binding the generated address to a public key to prove address ownership. IPv6 addresses are 128-bit IP addresses where the leftmost 64 bits form the \textit{subnet prefix} and the rightmost 64 bits form the \textit{interface identifier} \cite{HinIPv6}. The \textit{subnet prefix} is used to determine the host’s location in the Internet topology and the \textit{interface identifier} is used as an identity of the host.

A CGA is an IPv6 address whose \textit{interface identifier} is obtained by hashing the host’s public key and auxiliary parameters, together known as the CGA Parameters data structure. Any host can verify that a message comes from the claimed address by recomputing the address from the CGA Parameters data structure and verifying the attached signature, which must be valid for the public key.

The CGA Parameters data structure comprises a 128-bit randomly generated \textit{modifier} value, the 64-bit \textit{subnet prefix} of the address, an 8-bit \textit{collision count} value, the DER-encoded public key of the host, and a variable length \textit{extension field} value. The \textit{modifier} is used to strengthen the binding between the public key and the address and enhance privacy by adding randomness to the address. The \textit{collision count} can only take value 0, 1, or 2, and is used during address generation to recover from an address collision detected by Duplicate Address Detection (DAD) \cite{ThoSLAAC}. The \textit{extension field} can be used for additional data items. By default, it has length 0.

Each CGA also has a 3-bit security parameter \textit{sec} encoded in the three leftmost bits of its \textit{interface identifier}. The \textit{sec} parameter determines the strength of the binding between the public key and the address. It can have values from 0 (lowest security) to 7 (highest security). Hosts should select this value depending on their computational power. The higher the \textit{sec} value, the longer it takes to generate an address.

\begin{figure}[t]
    \centering
    \includegraphics[scale=0.22]{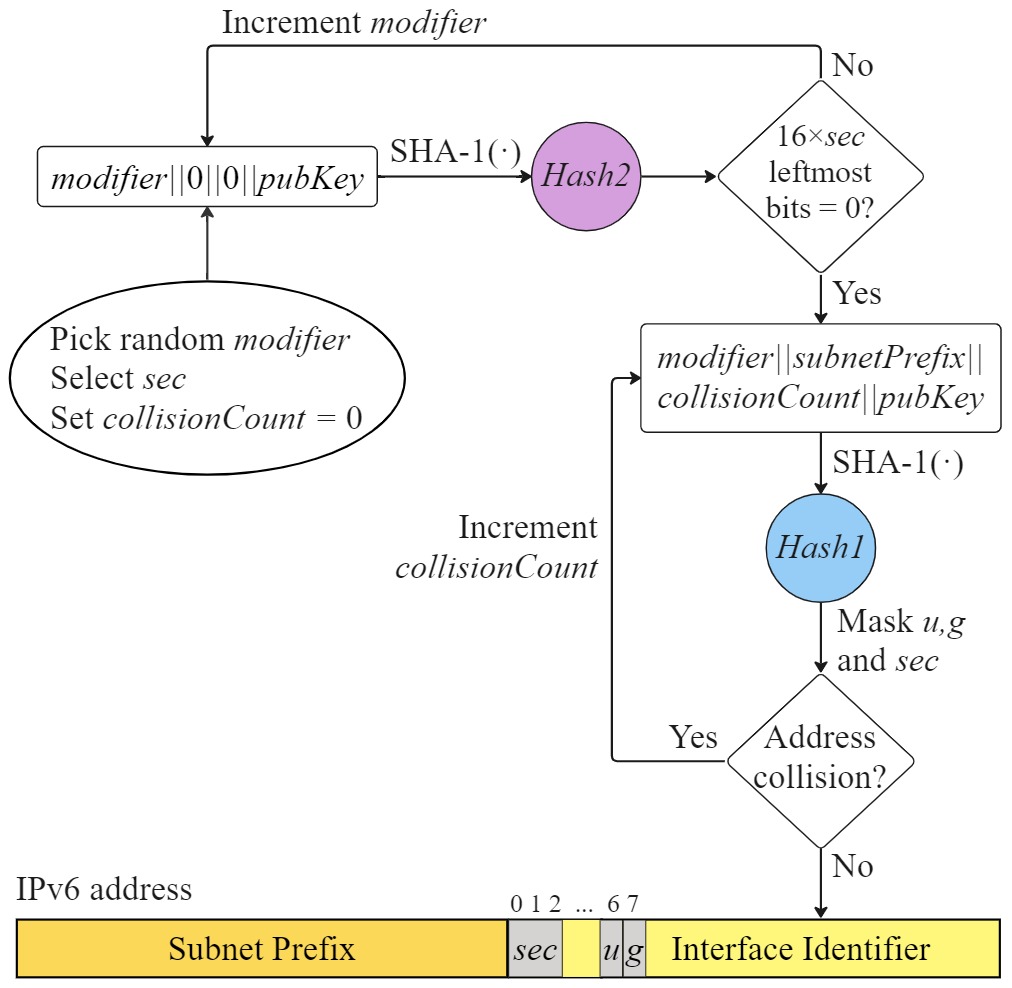}
    \caption{Detailed data flow of the CGA generation algorithm.}
    \label{fig:cga-generation}
\end{figure}

Fig.~\ref{fig:cga-generation} illustrates the CGA generation algorithm. It starts with the hash extension technique, in which hosts solve a partial hash inversion puzzle tied to their public key. Specifically, hosts iterate the \textit{modifier} until the $16 \times \textit{sec}$ leftmost bits of \textit{Hash2} are equal to zero. \textit{Hash2} contains the 112 leftmost bits of the hash digest computed over the CGA Parameters data structure with the \textit{subnet prefix} and \textit{collision count} set to zero. Then, \textit{Hash1} is computed by hashing the CGA Parameters data structure and taking the 64 leftmost bits of the hash digest. The \textit{interface identifier} of the address is derived from \textit{Hash1} by setting the \textit{u} and \textit{g} bits \cite{HinIPv6} (respectively 6\textsuperscript{th} and 7\textsuperscript{th} leftmost bits, starting from 0) to zero and encoding the \textit{sec} parameter in the three leftmost bits. Finally, the address is obtained by concatenating the \textit{subnet prefix} with the \textit{interface identifier}. If an address collision is detected by DAD, the \textit{collision count} is incremented by one and the \textit{Hash1} value is recomputed. After three collisions, the algorithm stops and an error is reported.

The original specification of CGA suggests the use of the SHA-1 algorithm to compute \textit{Hash1} and \textit{Hash2} \cite{AurCGA}. Due to the vulnerabilities found in SHA-1 \cite{WangSHA1}, propositions have been made to transition to more secure hash algorithms \cite{BagMultiHash}. For example, the SHA-256 hash algorithm is employed in several alternative CGA designs \cite{AlSCSCGA, CheECC, ShaCGAAnalysis}.

The CGA verification algorithm takes as input the CGA and its associated CGA Parameters data structure. The verification starts by checking that the \textit{collision count} is less than or equal to 2. Next, it checks that the \textit{subnet prefix} in the CGA Parameters data structure is the same as the one in the address. The verification then recomputes \textit{Hash1} and checks that it matches with the \textit{interface identifier} of the address. Finally, it recomputes \textit{Hash2} and verifies that the $16 \times \textit{sec}$ leftmost bits of the recomputed value are zero.

\subsection{CGA Analysis}

Spoofing a CGA implies finding a CGA Parameters data structure that comprises the attacker’s public key and successfully binds to the target CGA. The attacker can then use her private key to sign messages and pretend to be the legitimate owner of the address. In the following, we analyze the cost of this attack. We adapt the analysis made by Bos, Özen, and Hubaux \cite{BosCGA++} to the case where all three values of the \textit{collision count} can be iterated to spoof the CGA.

\begin{lemma}
Given a network, assume hosts generate IPv6 addresses using CGA with security parameter $0 \leq \textit{sec} \leq 7$. Then, the expected number of hash function evaluations required to spoof a specific address is:
\begin{equation}
T_{CGA} = \begin{cases}
  2^{59}  & \text{if } \textit{sec} = 0, \\
  2^{59} + (2^{16 \times \textit{sec}+59} / 3) & \text{if } \textit{sec} > 0.
\end{cases}
\end{equation}
\end{lemma}

\begin{proof}
Spoofing an address generated with CGA requires finding a valid CGA Parameters data structure that comprises the attacker’s public key and fulfills these conditions: (1) the leftmost $16 \times \textit{sec}$ bits of \textit{Hash2} are zero, and (2) the \textit{Hash1} value yields the target address. An attacker can proceed either by first satisfying condition (1) and then (2) or vice versa.

When starting with condition (1), the attacker is expected to perform $2^{16 \times \textit{sec}}$ hash function evaluations to find a suitable \textit{modifier}. For a fixed \textit{collision count}, the probability that condition (2) is satisfied is $2^{-59}$. Because the attacker can try the three different values of \textit{collision count} to satisfy condition (2), \textit{Hash2} is expected to be computed $2^{16 \times \textit{sec}+59} / 3$ times in total. Condition (2) is satisfied after computing \textit{Hash1} $2^{59}$ times on average. The total cost for spoofing, when starting with condition (1), thus becomes $2^{59} + (2^{16 \times \textit{sec}+59} / 3)$ hash function evaluations.

When starting with condition (2), the attacker iterates the inputs to \textit{Hash1} until condition (2) is satisfied. This is expected to require $2^{59}$ hash function evaluations. Next, the attacker computes \textit{Hash2} and verifies that condition (1) is satisfied, which happens with a probability $2^{-16 \times \textit{sec}}$. So, the attacker is expected to compute \textit{Hash1} $2^{16 \times \textit{sec}+59}$ times. Moreover, the attacker is expected to compute \textit{Hash2} $2^{16 \times \textit{sec}}$ times before condition (1) is satisfied. Therefore, the total cost for spoofing, when starting with condition (2), becomes $2^{16 \times \textit{sec}} + 2^{16 \times \textit{sec}+59}$ hash function evaluations.

For all \textit{sec} values with $1 \leq \textit{sec} \leq 7$, the cost of the attack is smaller when starting with condition (1) than with condition (2). Therefore, this attack has a cost $2^{59} + (2^{16 \times \textit{sec}+59} / 3)$ hash function evaluations. If $\textit{sec} = 0$, then the computation of \textit{Hash2} is skipped and condition (2) is satisfied after performing an average of $2^{59}$ hash function evaluations.
\end{proof}

The security of the binding between the public key and the address depends on the \textit{sec} parameter. Incrementing the \textit{sec} value by one increases the cost of spoofing attacks by a factor $2^{16}$. At the same time, it increases the cost of address generation by the same factor, thereby introducing a trade-off between security and performance. The cost of address verification is independent of the \textit{sec} parameter and is always two hash function evaluations.

Several studies found that generating addresses with a \textit{sec} value of 2 takes several minutes on standard devices, which is not an acceptable delay for generating addresses \cite{AlSTBCGA, BosCGA++, AlsSEND}. This implies that the \textit{sec} value cannot be larger than 1, which yields approximately 73 bits of security. For low-end devices, only a \textit{sec} with a value of zero is practical \cite{CheECC, QadCGAMobile}. This yields 59 bits of security which is far from being considered secure. These results suggest that generating secure CGAs on the fly is not practical for most devices.

\section{Bitcoin} \label{section:bitcoin}

The Bitcoin blockchain is a distributed ledger that acts as an immutable record of data \cite{NakBit}. As Bitcoin is public, anyone can publish data on the blockchain embedded in Bitcoin transactions. The Bitcoin network consists of nodes who are responsible for validating transactions and aggregating them into blocks. These blocks of transactions are appended to the blockchain via a stochastic process called mining. This process requires nodes to solve a partial hash inversion puzzle known as Proof of Work (PoW) \cite{JakPoW}.

A Bitcoin block contains an ordered list of transactions and a header which is of a fixed size 80 bytes. The block header contains multiple fields among which are the difficulty of the PoW of the block, a nonce value which can be iterated to solve the PoW, and a Merkle root obtained by hashing the ordered list of transactions into a Merkle tree \cite{Merkle}.

The PoW in Bitcoin consists of iterating the values in the block header until its double SHA-256 hash (SHA-256 applied twice) is below some target value. The target is derived from the difficulty value encoded in the block header. Nodes use a predefined algorithm to adjust the difficulty of the PoW to the computational power of the Bitcoin network. The algorithm ensures that the time required by the network to solve the PoW is 10 minutes on average.

A Bitcoin transaction contains virtually any number of inputs and outputs. Transaction outputs lock a certain number of bitcoins which can be unlocked by providing the correct data (e.g. a set of signatures) in the input of a new transaction. Outputs can also be used to inscribe data, such as hash values or text, onto the blockchain. Transactions include a fee that remunerates nodes for validating transactions, solving the PoW, and publishing new blocks.

Bitcoin relies on Merkle trees to commit to the transactions appearing in the block. The leaves of the Merkle tree consist of the transactions and each non-leaf node is labeled with the hash of the concatenation of its child nodes. The advantage of using Merkle trees is that it allows anyone to produce concise, unique, and easy-to-verify inclusion proofs of transactions. Given a transaction and its Merkle proof, a verifier can recompute the Merkle root and compare it with the Merkle root in the block header. If they match, the verifier can be assured that the transaction has been included in the block.

As of February 2023, the three main implementations of Bitcoin are Bitcoin Core (BTC) \cite{BTC}, Bitcoin Cash (BCH) \cite{BCH}, and Bitcoin SV (BSV) \cite{BSV}. Table I shows a comparative analysis of these three implementations. The transaction throughput refers to the number of transactions that can be added to the blockchain per second. Among these implementations, BSV offers the highest transaction throughput and cheapest transaction fees.

\section{Bitcoin-Certified Addresses (BCA)} \label{section:bca}
In this section, we present Bitcoin-Certified Addresses (BCA), an enhanced version of CGA whereby hosts delegate the work that secures the binding between the public key and the IPv6 address to Bitcoin nodes.

\subsection{Design Rationale}

BCAs are generated from a Bitcoin \textit{transaction} registering the public key of a host and the header of a Bitcoin block containing the \textit{transaction}. The search for a valid \textit{block header} performed by Bitcoin nodes during the PoW process is used to secure the binding between the public key and the address, similar to the search for a valid \textit{modifier} in CGA. Therefore, no computationally intensive operations have to be performed by hosts during address generation. Because the amount of work performed by Bitcoin nodes is typically much higher than what can be performed by the conventional machine of a host in CGA, the security of the binding is stronger in BCA than what can be achieved in practice in CGA.

% \begin{figure}[t]
%     \centering
%     \includegraphics[scale=0.2]{images/Merkle tree.jpg}
%     \caption{A block header contains the root of the Merkle tree of transactions. In this example, the Merkle proof of inclusion of the transaction \textit{Tx3} contains three hash values (in orange).}
%     \label{fig:merkle-tree}
% \end{figure}

BCA ensures that multiple addresses can be generated from a single \textit{transaction} and public key, thus avoiding the need to create and broadcast a new \textit{transaction} every time a new address is needed. This is achieved by creating a Merkle tree of \textit{modifier} values and including the root of the tree in the \textit{transaction} where the public key is registered. The \textit{modifier} values are then used as input to the BCA generation algorithm. They should be randomly generated to ensure that addresses generated from the same public key are un-linkable, thus protecting the privacy of hosts.

Each BCA is associated a BCA Parameters data structure, whose format is depicted in Table \ref{tab:bca-parameters-data-structure}. The BCA Parameters data structure is communicated to other hosts during address verification. To prevent an attacker from spamming hosts by sending very large (and potentially incorrect) proofs to be verified, we impose a limit $N_{max}$ to the number of \textit{modifier} values that can be committed in the \textit{transaction}. We also assume that the number of transactions appearing in a block does not exceed $M_{max}$, which implies that the \textit{Merkle proof of the transaction} should not contain more than $\log_2 (M_{max})$ hash values. We recommend using the values $N_{max} = 32$ and $M_{max} = 2^{28}$, which allows for more than 250 million transactions to be included in a block.

\begin{table}[t]
\caption{Comparison of BTC, BCH, and BSV}
\begin{center}
\begin{tabular}{l|c|c|c|l}
\cline{2-4}
                                                                                     & \textbf{\begin{tabular}[c]{@{}c@{}}Average\\ difficulty of \\ PoW (Feb. \\ 2023) \cite{BitDiff}\end{tabular}} & \textbf{\begin{tabular}[c]{@{}c@{}}Transaction fees\\ range (Feb. 2023) \\ \cite{TxFee, TxFeeBSV} \end{tabular}} & \textbf{\begin{tabular}[c]{@{}c@{}}Max. recorded\\ throughput (up\\ to Feb. 2023) \\ \cite{TxRateBTC, TxRateBCH, TxRateBSV} \end{tabular}} &  \\ \cline{1-4}
\multicolumn{1}{|l|}{\begin{tabular}[c]{@{}l@{}}Bitcoin\\ Core\\ (BTC)\end{tabular}} & $40 \times 10^{12}$                                                                                   & \$1.3--2.5                                                                                                        & \textless{10} tx/sec                                                                                                        &  \\ \cline{1-4}
\multicolumn{1}{|l|}{\begin{tabular}[c]{@{}l@{}}Bitcoin\\ Cash\\ (BCH)\end{tabular}} & $225 \times 10^{9}$                                                                                  & \$0.002--0.007                                                                                                        & \textless{200} tx/sec                                                                                                          &  \\ \cline{1-4}
\multicolumn{1}{|l|}{\begin{tabular}[c]{@{}l@{}}Bitcoin\\ SV\\ (BSV)\end{tabular}}   & $83 \times 10^{9}$                                                                                   & \$0.00005--0.002                                                                                                        & 18,606 tx/sec                                                                                                                   &  \\ \cline{1-4}
\end{tabular}
\label{tab:btc-bch-bsv}
\end{center}
\end{table}

Like CGA, each BCA has a security parameter \textit{sec} encoded in the three leftmost bits of its \textit{interface identifier}. This parameter is derived from the \textit{difficulty} of the PoW computed by Bitcoin nodes and determines the strength of the BCA against spoofing attacks.

\subsection{BCA Specification}

\begin{table}[t]
\caption{BCA Parameters data structure}
\begin{center}
\renewcommand{\arraystretch}{1.25}
\begin{tabular}{|c|c|}
\hline
\textit{modifier}                    & 16 bytes                         \\ \hline
\textit{public key}                  & variable length                  \\ \hline
\textit{transaction}                 & variable length                  \\ \hline
\textit{block header}                & 80 bytes                         \\ \hline
\textit{subnet prefix}               & 8 bytes                          \\ \hline
\textit{collision count}             & 1 byte                           \\ \hline
\textit{Merkle proof of transaction} & max. 896 bytes (with $M_{max} = 2^{28}$) \\ \hline
\textit{Merkle proof of modifier}    & max. 160 bytes (with $N_{max} = 32$)    \\ \hline
\end{tabular}
\label{tab:bca-parameters-data-structure}
\end{center}
\end{table}

\noindent \textbf{Public key registration.} In order to generate BCAs, hosts must register their public key on the Bitcoin blockchain. For this, they create and broadcast a Bitcoin \textit{transaction} whose data payload includes the hash of their public key. They also generate a list of $N$ random 128-bit \textit{modifier} values, with $N \le N_{max}$, and include the root of the associated Merkle tree in the \textit{transaction}. Hosts must store these \textit{modifier} values and the associated Merkle tree in memory. Once the \textit{transaction} is included in a block, hosts fetch and verify the \textit{Merkle proof} of inclusion of the \textit{transaction} in the block and store it in memory. They also fetch the \textit{block header} of the block where the \textit{transaction} is included and store it in memory.

The \textit{sec} parameter is derived from the \textit{difficulty} of the PoW encoded in the \textit{block header} as $\textit{sec} = \lfloor \log_2 (\textit{difficulty}) / 16 \rfloor$. This implies that for a given \textit{difficulty}, the double SHA-256 hash of the \textit{block header} found by Bitcoin nodes has its $16 \times (\textit{sec} + 2)$ leftmost bits equal to zero. The minimum \textit{sec} value is 0 and is obtained when the \textit{difficulty} of the PoW is 1, its minimum value. This corresponds to a PoW that requires finding a hash digest whose leftmost 32 bits are zero. The maximum \textit{sec} value is 7, which requires finding a hash digest whose leftmost 144 bits are zero.

\noindent \textbf{BCA generation.} We define \textit{modifier\textsubscript{i}} as the \textit{i}-th modifier in the list of generated \textit{modifier} values. The index \textit{i} is initialized to zero and is incremented by one every time a new address in the subnet is generated. The BCA generation algorithm takes as input the \textit{modifier\textsubscript{i}}, the \textit{block header}, the \textit{subnet prefix}, and the \textit{transaction}, and works as follows.

\begin{enumerate}

    \item Set \textit{collision count} to zero.
    \item Hash the concatenation of the \textit{modifier\textsubscript{i}}, the \textit{block header}, the \textit{subnet prefix}, the \textit{collision count}, and the \textit{transaction}, and take the leftmost 64 bits of the resulting hash value. The result is \textit{Hash1}.
    \item Construct the \textit{interface identifier} from \textit{Hash1} by writing the \textit{sec} value into the three leftmost bits and setting the \textit{u} and \textit{g} bits to zero.
    \item Concatenate the \textit{subnet prefix} and the \textit{interface identifier} to form a 128-bit IPv6 address.
    \item Perform DAD. If an address collision is detected, increment the \textit{collision count} by one and go back to step 2. After three collisions, stop and report an error.

\end{enumerate}

\begin{figure}[t]
    \centering
    \includegraphics[scale=0.25]{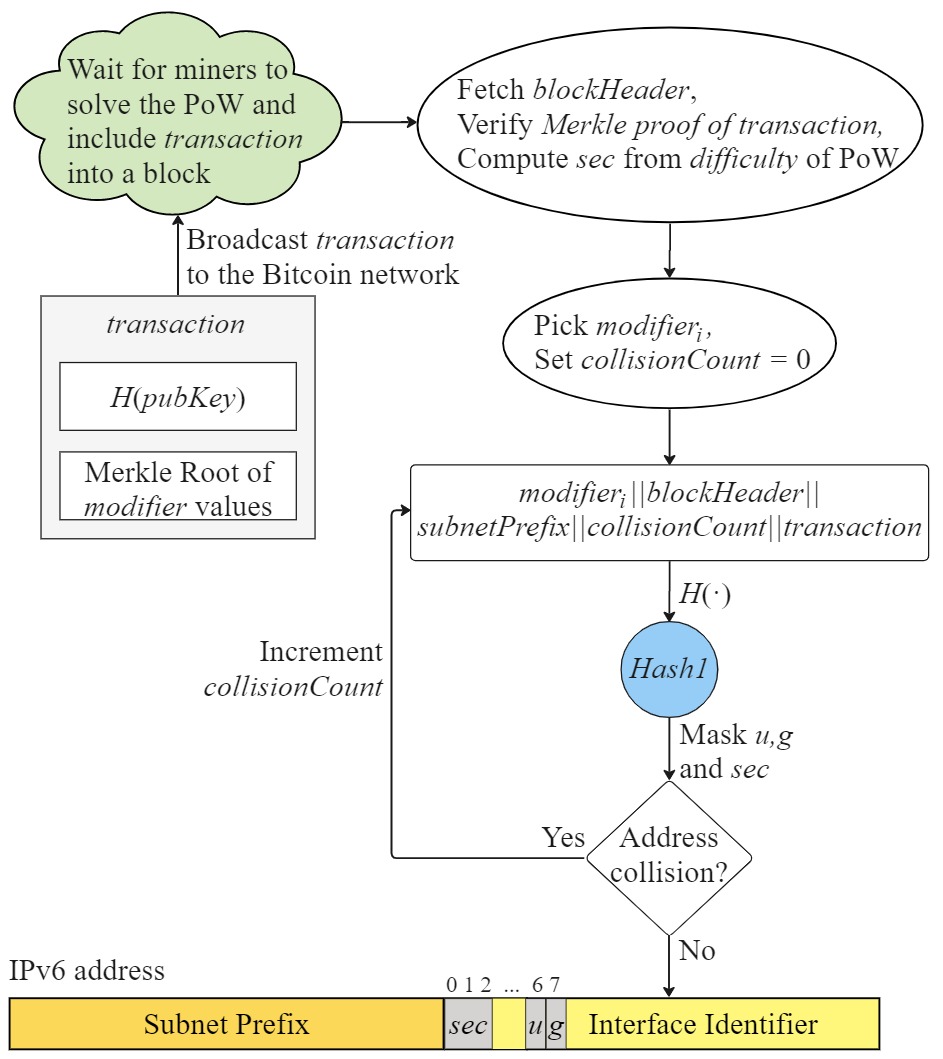}
    \caption{Detailed data flow of the public key registration process and BCA generation algorithm.}
    \label{fig:bca-generation}
\end{figure}

The output is a new BCA and its associated BCA Parameters data structure. Fig. \ref{fig:bca-generation} summarizes the public key registration process and BCA generation algorithm.

\noindent \textbf{BCA verification.} The BCA verification algorithm takes as input the BCA and its associated BCA Parameters data structure, and checks that the following conditions hold:

\begin{enumerate}
    \item The \textit{collision count} is equal to 0, 1, or 2.
    \item The \textit{subnet prefix} in the BCA Parameters data structure is equal to the \textit{subnet prefix} of the address.
    \item The hash of the public key is equal to the hashed public key included in the data payload of the \textit{transaction}.
    \item The format of the \textit{block header} is valid.
    \item The \textit{Merkle proof} of inclusion of the \textit{transaction} in the block is valid and the height of the corresponding Merkle tree is less than or equal to $\log_2(M_{max})$ (there are less than $M_{max}$ transactions in the block).
    \item The \textit{Merkle proof} of inclusion of \textit{modifier\textsubscript{i}} in the \textit{transaction} is valid and the height of the corresponding Merkle tree is less than or equal to $\log_2(N_{max})$ (there are less than $N_{max}$ \textit{modifier} values committed in the \textit{transaction}).
    \item Extract the \textit{sec} parameter from the three leftmost bits of the \textit{interface identifier} of the address. The leftmost $16 \times (\textit{sec} + 2)$ bits of the double SHA-256 hash of the \textit{block header} are zero.
    \item Compute \textit{Hash1} from the BCA Parameters data structure. \textit{Hash1} is equal to the \textit{interface identifier} of the address. Differences in the \textit{u}, \textit{g}, and \textit{sec} bits are ignored.
\end{enumerate}

If each condition is met, then the binding between the public key in the BCA Parameters data structure and the address is valid.

\subsection{BCA Analysis}

In the following, we analyze the resistance of BCA against spoofing attacks and compare BCA to CGA.

\begin{lemma}
Given a network, assume hosts generate IPv6 addresses using BCA with security parameter $0 \leq \textit{sec} \leq 7$. Assume there is a limit $N_{max}$ on the number of modifier values committed in the transaction and a limit $M_{max}$ on the number of transactions contained in a Bitcoin block. Then, the expected number of hash function evaluations required to spoof a specific address is:
\begin{equation}
T_{BCA} = 2^{59} + \frac{2^{16 \times \textit{sec}+92}}{3 \times N_{max} \times M_{max}}.\label{eq:bca}
\end{equation}

% \begin{equation}
% T_{BCA} = 2^{59} + (2^{16 \times sec+92} / (3 \times N_{max} \times M_{max})).
% \end{equation}
\end{lemma}

\begin{proof}
Spoofing an address generated with BCA requires constructing a dummy Bitcoin block that contains a transaction registering the attacker’s public key and that fulfills the following two conditions: (1) the leftmost $16 \times (\textit{sec} + 2)$ bits of the double SHA-256 hash of the \textit{block header} are zero, and (2) the \textit{Hash1} value yields the target address. The attack can be conducted in two ways: by first satisfying condition (1) and then (2), or vice versa. Note that the block does not need to be published to the blockchain for the attack to be successful. The attacker can include any number of transactions $M$ registering her public key, such that $M \leq M_{max}$. Additionally, the attacker can commit to any number of \textit{modifier} values $N \leq N_{max}$ in the transactions.

When starting with condition (1), the attacker is expected to perform $2^{16 \times (\textit{sec}+2)+1}$ hash function evaluations to find a suitable \textit{block header}. Because the probability that one of the $N$ \textit{modifier} values, one of the $M$ transactions, and one of the three \textit{collision count} values satisfy condition (2) is $2^{-59} \times 3 \times N \times M$, this process needs to be repeated $2^{59} / (3 \times N \times M)$ times on average. This suggests that the attack is the most efficient when $N$ and $M$ are maximal, that is the block created by the attacker contains $M_{max}$ transactions and each transaction commits to $N_{max}$ \textit{modifier} values. Condition (2) is satisfied after being tested $2^{59}$ times on average. The total cost for spoofing, when starting with condition (1), thus becomes $2^{59} + \frac{2^{16 \times \textit{sec}+92}}{3 \times N_{max} \times M_{max}}$ hash function evaluations.

When starting with condition (2), the attacker iterates the parameters of the \textit{block header} until condition (2) is satisfied. This is expected to require $2^{59}$ hash function evaluations. Next, the attacker computes the double SHA-256 hash of the \textit{block header} to verify that condition (1) is satisfied, which happens with a probability $2^{-16 \times (\textit{sec}+2)}$. On average, the double SHA-256 hash of $2^{16 \times (\textit{sec}+2)}$ different block headers have to be computed before condition (1) is satisfied. Therefore, the total cost for spoofing, when starting with condition (2), becomes $2^{16 \times (\textit{sec}+2)+1} + 2^{16 \times (\textit{sec}+2)+59}$ hash function evaluations.

For all \textit{sec} values with $0 \leq \textit{sec} \leq 7$, the cost of the attack is smaller when starting with condition (1) than with condition (2). Therefore, creating a BCA Parameters data structure that binds the attacker’s public key to some target address requires on average $2^{59} + \frac{2^{16 \times \textit{sec}+92}}{3 \times N_{max} \times M_{max}}$ hash function evaluations.

\end{proof}

The resistance of BCA against spoofing attacks is higher than what can be achieved in practice with CGA. By using the values $N_{max} = 32$ and $M_{max} = 2^{28}$ in \eqref{eq:bca}, the cost of spoofing attacks becomes $T_{BCA} = 2^{59} + (2^{16 \times \textit{sec}+59} / 3)$ hash function evaluations. As of February 2023, the difficulty of the PoW in the three main implementations of Bitcoin (BTC, BCH, and BSV) is such that the \textit{sec} value is 2 (cf. Table \ref{tab:btc-bch-bsv}), yielding a security of $\sim$89 bits. As a comparison, the maximum security that can be attained in CGA on a standard device with a reasonable address generation time is by using \textit{sec} equal to 1 \cite{AlSTBCGA, BosCGA++, AlsSEND}, which yields $\sim$73 bits of security.

Once the public key is registered on the blockchain, generating BCAs is fast and requires a single hash function evaluation. Like CGA, mobile hosts changing subnet can quickly obtain a new address by recomputing \textit{Hash1} with the new \textit{subnet prefix}. Verifying an address generated with BCA implies verifying two Merkle proofs. This is a slightly higher requirement than CGA, which demands two hash function evaluations, but remains still lightweight. Note that the verification does not require any connection to the Bitcoin network.

BCA offers better privacy than CGA. In CGA, changing an address implies redoing all the expensive work required during CGA generation, which disincentivizes hosts from changing their address frequently. In BCA, changing an address can efficiently be done by selecting a different \textit{modifier} that was committed in the \textit{transaction} and recomputing \textit{Hash1} with the new value. By committing to $N_{max} = 32$ \textit{modifier} values in the \textit{transaction}, hosts can change their address every day for a month in the same subnet without having to broadcast a new \textit{transaction} to the Bitcoin network. When hosts change subnet, they may reuse the old \textit{modifier} values to generate new addresses. This further minimizes the interaction needed with the Bitcoin network but may make it easier for an observer to link addresses with each other.

In comparison to CGA, BCA incurs extra storage overheads, but they do not exceed a few kilobytes. First, hosts have to store a maximum of $N_{max}$ \textit{modifier} values in memory and the associated Merkle tree. Assuming that 32 128-bit \textit{modifier} values are generated and that the nodes of the Merkle tree are of size 256 bits, then around 2.5 extra kilobytes of data have to be stored. Moreover, hosts have to store the \textit{Merkle proof of the transaction}. Assuming that Bitcoin blocks do not contain more than $2^{28}$ transactions, then the size of the \textit{Merkle proof of the transaction} to be stored does not exceed 1 kilobyte. 

\subsection{Deployment Considerations}

It is possible to combine BCA and CGA with one another to improve the usability of both techniques. The prerequisite for generating BCAs is to register a public key on the blockchain. Because a transaction is included in the blockchain after 10 minutes on average, hosts must perform the public key registration process in advance, rather than when a new address is needed. For hosts who have not yet registered their public key and would like to quickly obtain a new address, it might be desirable to use CGA with a low \textit{sec} value. By using a low \textit{sec} value, CGA generation is fast but the resistance of the address against spoofing attacks is low. This approach is acceptable if the address is a temporary one. Whenever a connection with the Bitcoin network is established, hosts should register their public key on the blockchain and generate addresses using BCA that are more resistant to spoofing attacks.

BCA requires the Bitcoin network to be able to process a large number of transactions. The transaction fees should also be as low as possible to minimize costs for hosts to register their public key on the blockchain. In 2023, it is estimated that the number of networked devices is roughly 29.3 billion \cite{Cisco}. Assuming all devices generate IPv6 addresses with BCA and they register their public key on the blockchain once a month, an average of \textgreater10,000 transactions per second would be produced on the network. According to Table \ref{tab:btc-bch-bsv}, Bitcoin SV provides the highest transaction throughput and lowest transaction fees, making it the most adapted Bitcoin implementation to integrate in BCA.

\section{Implementation and Evaluation} \label{section:implementation}

In this section, we describe our BCA implementation, evaluate its security and cost, and compare its performances to that of CGA.

For the purpose of our implementation, we employ Elliptic Curve Cryptography with the secp256k1 curve to generate public keys. The hash algorithm used is SHA-256. The parameters for the maximum number of \textit{modifier} values $N_{max}$ and the maximum number of transactions in a block $M_{max}$ are taken to be $N_{max} = 32$ and $M_{max} = 2^{28}$.

The public key registration process is performed on the Bitcoin SV blockchain. We start by randomly generating a public key and 32 \textit{modifier} values. The hash of the public key and the Merkle root of the \textit{modifier} values are included in a Bitcoin transaction \cite{WhatsOnChain}. The size of the resulting transaction is 341 bytes. At the time of writing, this requires paying only 0.00004 cents of USD in transaction fees, which is negligible for any user. The difficulty value of the block in which the transaction is included is $68.271 \times 10^{9}$. The resulting \textit{sec} value is 2, which according to \eqref{eq:bca} yields $\sim$89 bits of security.

We implemented the address generation algorithm of BCA and CGA in Python. The hashing part is implemented in C++ using the OpenSSL library \cite{OpenSSL}. We set the \textit{sec} value in CGA to 2 to evaluate both techniques with the same level of security. Table \ref{tab:eval-cga-bca} shows the result of our evaluation. The tests were performed on an Intel Core i7-1165G7 CPU running at 2.80 GHz on a single thread. As the results show, CGA generation took on average 696.15 s of CPU time to complete, with 32 tests performed. The best case required around 167 s, while the worst case more than 20 minutes. This shows the high computational cost and variance of CGA generation, which is undesirable for users. On the other hand, BCA generation took only 0.00096 ms on average.

\begin{table}[t]
\caption{CGA and BCA Generation Results, $\textit{sec}=2$}
\begin{center}
\renewcommand{\arraystretch}{1.25}

\begin{tabular}{l|l|l|}
\cline{2-3}
                                                           & CGA & BCA \\ \hline
\multicolumn{1}{|l|}{Expected number of hash evaluations}            & $2^{32}$ & 1    \\ \hline
\multicolumn{1}{|l|}{Hash input len (bits)} & 904    & 3,568    \\ \hline
\multicolumn{1}{|l|}{Sample size}                          & 32   & 1,000    \\ \hline
\multicolumn{1}{|l|}{Average address generation time}         & 696.15 s    & 0.00096 ms   \\ \hline
\end{tabular}
\label{tab:eval-cga-bca}
\end{center}
\end{table}

\section{Conclusion} \label{section:conclusion}

In this paper, we presented Bitcoin-Certified Addresses (BCA), a new technique to bind IPv6 addresses to public keys to prove address ownership. BCA solves the inefficiency issues of CGA by delegating the expensive work needed to secure the binding to the highly resilient network of Bitcoin nodes. As opposed to CGA, BCA does not trade security for performance, it offers both. Once the public key of hosts is registered on the blockchain, it is easy to generate and change addresses which improves the privacy of hosts compared to CGA. To demonstrate the usability and efficiency of BCA, we offered an implementation relying on the Bitcoin SV network. 
% [Add conclusion]

Because of its low computational constraint, good security guarantees, and lack of reliance on a trusted authority or security infrastructure, we believe that BCA is a promising technique for protecting IPv6 networks against address spoofing, especially in environments where devices have limited computational power such as in Internet of Things (IoT) networks. In the future, we should ensure that devices can seamlessly interact with the Bitcoin network to register their public key and use BCA to generate IPv6 addresses. This implies developing the tools and infrastructure that make funding and broadcasting transactions as easy as possible. Finally, this requires continuing to improve the capacity of the Bitcoin network so that it can scale with the increasing number of transactions.

\section*{Acknowledgements}

The author thanks Michaella Pettit, Wei Zhang, Luigi Lunardon, Alessio Pagani, and Enrique Larraia for their useful comments on the paper. The author also thanks John Murphy for implementing the BCA and CGA techniques.

\end{document}